\documentstyle[amssymb,aps]{revtex}
\begin{document}
\title{Stationary states of \ Jaynes-Cummings model with atomic center-of-mass
quantum motion: direct comparison of standing-wave and
counterpropagating-waves cases.}
\author{A. Zh. Muradyan$^{1,2}$, G. A. Muradyan$^{1}$}
\address{$^1$Department of Physics, Yerevan State University, 1 Alex Manukian,\\
Yerevan 375049 Armenia; \\
$^2$Engineering Center of Armenian National Academy of Sciences,\\
Ashtarak-2, 378410 Armenia; \\
E-mail: muradyan@server.physdep.r.am}
\maketitle

\begin{abstract}
The eigenstate problem of the Jaynes-Cummings model on the basis of complete
Hamiltonian, including the center-of -mass kinetic energy operator, is
treated. \ The energy spectrum and wave functions in standing-wave (SW)- and
counterpropagating waves (CPW)- cases are calculated and compared with each
other. \ It is shown that in CPW-case i) the atomic momentum distribution is
asymmetric and somewhat narrower in general; ii) the concept of
quasimomentum is not applicable and instead the ordinary momentum concerns
the problem; iii) atomic and photonic state distributions are
self-consistent, and, in consequence iiii) mean number of photons in the
counterpropagating traveling waves and mean atomic momentum match. Explicit
analytic expressions for energy eigenvalues and eigenfunctions are found in
Tavis -Cummings-type approximation [Phys. Rev. 170, 379(1968)] and is
pointed, that it implies only the bounded-like states for atomic
center-of-mass motion. \ It is also shown that if the recoil energy is taken
into account, the Doppleron resonance is split into two branches, one of
which diverges to Bragg-like resonance in the high-order range.
\end{abstract}

\section{Introduction}

The key scheme of cavity quantum electrodynamics (QED)\cite{1}, modern atom
optics and interferometry \cite{2} is the resonant interaction of an atom
with cavity fields, created as standing or counterpropagating waves. \ The
first type (with respective quantization on the SW basis) is attained in two
plane-parallel-mirror cavities (in short, cavity), while the second type
(with respective CPW-quantization) is attained in three or more mirror
cavities (ring cavities). \ It is well known, in addition, that for
classical picture of fields, these two representations are equivalent in the
sense, that the SW can always be presented as a superposition of two
counterpropagating travelling waves. \ In the quantum theory, nevertheless,
they are divers\cite{3}, including the Hamiltonians of interaction. \ Hence
it needs to be ascertained: are the diversities only quantitative or
qualitative too, and how much are they for these or that circumstances and
processes. Note, that it presents not only an academic interest, since in
the microcavities ($V\precsim 10^{-3}cm^{3}$) several dozens of photons can
induce strong optical nonlinearities for dipole-allowed optical transitions.

As far as we know, elucidation of physical aspects of mentioned incongruity
had been done in paper \cite{4}, taking as an example the process of
near-resonant coherent diffraction of atomic matter-waves by a
space-periodic laser radiation field, known as near-resonant Kapitza-Dirac
diffraction\cite{5}. \ By means of numerical solution of the master
equations for atomic probability amplitudes, the existence of evident
differences between SW and CPW diffraction patterns was shown , in strictly
quantum domain of cavity fields (the mean number of photons in cavities was
chosen one). \ The physical reason of the difference also was presented. \
It is the behavior of momentum conservation law in relevant ''atom + field
'' systems. \ This general principle ceases to apply in SW-case, because the
atom changes its own momentum due to the interaction, while the SW-field
does not. \ In contrary, in CPW-case both, atomic and field subsystems,
change the momenta equally and in opposite directions, thereby conserving
the systems total momentum. \ Already on the base of this sole difference
can be anticipated a ''soffer'' diffraction for CPW-case with a respectively
narrower distribution of states in the momentum space.

In this paper are considered the stationary states of the system ''two-level
atom + quantized field'' in both SW and CPW-cases and the qualitative and
quantitative differences between them are carried out in the range of
quantum optics (the number of photons should be less than several tens). \
To this end the simplest theoretical model, the Jaynes-Cummings model (JCM) 
\cite{6}, including the quantized atomic center-of-mass motion \cite{7},
will be used. The spontaneous emission and other incoherent processes, as
usual for this model, are omitted. \ 

To make possible the direct comparison of these two concepts we, at first,
represent detailed theory for stationary states of the systems at hand. \
Our representation for SW-case somewhat coincides with the picture in \cite
{8}. \ The preliminary analysis of the states and the comparison of the SW
and CPW-cases are based on numerical solutions of the exact, as well as of
the off-resonant approximate equations of atomic probability amplitudes in
momentum space. An additional approximate analytical solution of the problem
allows us to find out the parameters (terms), determining the size of
nonequivalency between the SW and CPW-spectrums and eigenfunctions in
explicit form. \ The latter solutions pertain, however,only to bounded
(wells' inner) atomic states.

The paper is organized as follows. Section II represents the basic model and
the exact set of equations. \ Here we also give some numerical simulations
of the problem, implemented for the sodium atom with $3S_{1/2}-3P_{3/2}$
main transition. \ Section III represents the off-resonance and mentioned
analytical approximations. \ In Section IV we examine the Bragg and
Doppleron resonances, including the recoil energy terms. \ In Section V we
consider the mechanisms responsible for the formation of stationary states
from free definite-momentum ones. \ Conclusions and some short remarks about
the subject are given in Section VI.

\section{Basic Model Theory}

We consider a two level atom of mass $M$ and optical transition frequency $%
\omega _{0}$, interacting with a quantized plane monochromatic field of
frequency $\omega $. \ We will restrict ourselves with two important cases
of standing and counterpropagating waves. \ The longitudinal components $%
p_{x},p_{y}$ of the atomic center-of-mass (c.m.) momentum remain unchanged
for perfectly plane fields, so that only the transverse atomic momentum $%
p_{z}=p$ needs to be considered. \ Since the incoherent processes are not
included, the system is governed by the Schrodinger equation.

Let us first consider the CPW-case. \ The Hamiltonian in dipole and
rotating-wave approximations is given by \cite{4} 
\[
H=\frac{1}{2M}\widehat{P}^{2}+\frac{1}{2}\left( 1+\sigma _{3}\right) \hbar
\omega _{0}+\hbar \omega \left( a_{1}^{+}a_{1}+a_{2}^{+}a_{2}\right) 
\]

\begin{equation}
+\frac{1}{2^{3/2}}\hbar \Omega _{0}\left[ \sigma _{+}\left( a_{1}\exp
(ikz)+a_{2}\exp (-ikz)\right) +\sigma _{-}\left( a_{1}^{+}\exp
(ikz)+a_{2}^{+}\exp (-ikz)\right) \right] ,  \eqnum{1}  \label{1}
\end{equation}
where $\widehat{\overline{P}}=-i\hbar d/dz$ is atomic momentum operator
(1D), $\sigma _{3},\sigma _{+}$ and $\sigma _{-}$ are usual pseudospin
atomic (Pauli) operators, $\widehat{a}_{i}$ and $\widehat{a}_{i}^{+}$ $%
(i=1,2)$ are annihilation and creation operators for two running wave modes, 
$k=\omega /c$. The factor $\Omega _{0}$ in coupling constant is the vacuum
Rabi frequency, \ and is connected with the atomic transition dipole moment $%
d$ by relation $\Omega _{0}/2^{3/2}=-d(2\pi \omega /\hbar V)$, where $V$ is
the normalization (cavity) volume. \ The coherent binding between the
internal and c.m. variables is by the Rabi frequency $\Omega _{0}$, too.

The system under consideration has four degrees of freedom: \ two per atomic
internal and 1D c.m. motions, and two per counterpropagating travelling
waves. \ Three operators, forming together with the Hamiltonian the complete
set of mutually commutative operators, are known and are the following: the
excitation number operator 
\begin{equation}
\widehat{N}=\frac{1}{2}\left( 1+\sigma _{3}\right)
+a_{1}^{+}a_{1}+a_{2}^{+}a_{2},  \eqnum{2}  \label{2}
\end{equation}
the total momentum operator 
\begin{equation}
\widehat{P}=\widehat{p}+\hbar k\left( a_{1}^{+}a_{1}-a_{2}^{+}a_{2}\right) ,
\eqnum{3}  \label{3}
\end{equation}
and the operator 
\begin{equation}
\widehat{T}=\sigma _{3}\exp 
%TCIMACRO{\QOVERD( ) {i\pi \widehat{p}}{\hbar k}}%
%BeginExpansion
{i\pi \widehat{p} \overwithdelims() \hbar k}%
%EndExpansion
,  \eqnum{4}  \label{4}
\end{equation}
which combines translational $\widehat{p}$ and dipole inversing $\sigma _{3}$
operators . \ Therefore $\widehat{H},\widehat{N},\widehat{P}$ and $\widehat{T%
}$ have compatible eigenvalues and a common system of eigenfunctions. \ Only
these states, being the basic, should be considered thereafter in this
paper. \ Any other state, of course in principle, can be presented via these
eigenstates. \ Note, that sometimes the operator $\widehat{I}=\exp (i2\pi 
\widehat{p}/\hbar k)$ is used in related problems. \ However, $\widehat{I}=%
\widehat{T}^{2}$and can not be added into the set of mentioned operators or
replace the operator $\widehat{T}$ there.

Out of a desire to better understand the future solutions and their
connection with free-state values, which present a definite interest too, it
may be worth to emphasize here that all three $\widehat{N},\widehat{P}$ and $%
\widehat{T}$ operators don't contain the interaction parameter $\Omega _{0}$%
. \ They are interaction independent. \ Their eigenvalues, denoted by $N,P$
and $T$ respectively, also are interaction independent. \ Therefore, any
intermediate or final state, created from the initial one due to interaction
should possess the same values of $N,P$ and $T$ as the free initial state. \
These ones, of course, don't need to be stationary in general, but may be
such, if the interaction is adiabatic\cite{9}. A sufficient condition for
adiabatic following is the exceeding of the interaction switching on and
switching off times ($\tau _{sw}$), the inverse quantity of the system's
smallest characteristic frequency ($\Omega _{\min }^{-1}$). \ In the system
at hand, as a such frequency appears the one-photon recoil frequency $\Omega
_{r}=\hbar k^{2}/2M.$ \ So for the interaction times essentially exceeding
the recoil time $\Omega _{r}^{-1}$, the atomic state evaluation is adiabatic
with respect to both, internal and c.m., motions of the atom \cite{10}. \
Failure of the adiabatic condition implies, respectively, that the
population leaks into the immediate neighboring states, first of all into
the translational states. \ Applying the general remarks to the system under
consideration, we arrive to the possibility to treat the stationary states
as attained adiabatically from some free initial state and then replace the
set of $N,P$ and $T$ quantities by a new set of three related parameters of
this initial state: atomic momentum $p$ and the photon numbers $n_{1}$and $%
n_{2}$ in counterpropagating waves. \ One must keep in mind, however, that
the mentioned replacement, which will be embodied soon, is substantial for
adiabatically treated stationary states and is not obligatory even in this
case.

Dirac notations $\left| l\right\rangle _{1\text{ }}$and $\left|
m\right\rangle _{2\text{ }}$are used for quantized travelling-wave states
(number states). \ Atomic internal ground and excited states are specified
by $\left( 
\begin{array}{c}
0 \\ 
1
\end{array}
\right) $ and $\left( 
\begin{array}{c}
1 \\ 
0
\end{array}
\right) $ matrixes respectively, the c.m. states- by the function $\exp
(ipz/\hbar ).$ Hilbert space of the ''atom+field'' system is the direct sum
of Hilbert spaces of atomic c.m. and internal motions, as well as of the
field modes. \ Thereby, the nondegenerate wave function of the free system
can be written as 
\begin{equation}
\Psi _{free}=A\varphi _{j}\exp 
%TCIMACRO{\QOVERD( ) {ipz}{\hbar }}%
%BeginExpansion
{ipz \overwithdelims() \hbar }%
%EndExpansion
\left| l\right\rangle _{1}\left| m\right\rangle _{2},  \eqnum{5}  \label{5}
\end{equation}
where $A$ is the normalization constant, $j=g,e;\varphi _{g,e}=\left( 
\begin{array}{c}
0 \\ 
1
\end{array}
\right) ,\left( 
\begin{array}{c}
1 \\ 
0
\end{array}
\right) ;$ $-\infty \leq p\leq \infty $ and $l,m=0,1,2,\cdot \cdot \cdot $
in general. \ The seeking wave functions of the interacting system can be
written as a superposition of (\ref{5})-type terms.

The general form of a wave function with a definite $N$ ($\widehat{N}\Psi
=N\Psi $) is 
\[
\Psi =\left( 
\begin{array}{c}
0 \\ 
1
\end{array}
\right) \sum_{r=0}^{N}\int a_{r}(q,t)\left| r\right\rangle _{1}\left|
N-r\right\rangle _{2}\exp 
%TCIMACRO{\QOVERD( ) {iqz}{\hbar }}%
%BeginExpansion
{iqz \overwithdelims() \hbar }%
%EndExpansion
dq 
\]

\begin{equation}
+\left( 
\begin{array}{c}
1 \\ 
0
\end{array}
\right) \sum_{r=0}^{N-1}\int b_{r}(q,t)\left| r\right\rangle _{1}\left|
N-1-r\right\rangle _{2}\exp 
%TCIMACRO{\QOVERD( ) {iqz}{\hbar }}%
%BeginExpansion
{iqz \overwithdelims() \hbar }%
%EndExpansion
dq,  \eqnum{6}  \label{6}
\end{equation}
where $a_{r}$ and $b_{r}$ are arbitrary coefficients yet. Admissible
solutions also should be normalized according to $\left\langle \Psi \mid
\Psi \right\rangle $ $=1.$ \ Note, that the excitation number $N$ can be
represented by means of the numbers $n_{1}$ and $n_{2}$ (see the remark
above ): 
\begin{equation}
N=\left\{ 
\begin{array}{c}
n_{1}+n_{2}\text{ for free ground level atom,} \\ 
n_{1}+n_{2}+1\text{ for free excited level atom. }
\end{array}
\right\} ,  \eqnum{7}  \label{7}
\end{equation}
validity of which can be directly checked with the help of wave function (%
\ref{6}).

For the next step we demand from $\Psi $ to be the eigenfunction of total
momentum operator $\widehat{P}$: 
\begin{equation}
\widehat{P}\Psi =P\Psi .  \eqnum{8}  \label{8}
\end{equation}
Inserting (\ref{3}) and (\ref{6}) into (\ref{8}) we arrive to equations 
\begin{equation}
a_{r}(q,t)=a_{r}(t)\delta \left( q+(2r-N)\hbar k-P\right) ,  \eqnum{9a}
\label{9a}
\end{equation}

\begin{equation}
b_{r}(q,t)=b_{r}(t)\delta \left( q+(2r+1-N)\hbar k-P\right) ,  \eqnum{9b}
\label{9b}
\end{equation}
where $\delta (x)$ is the Dirac delta - function and $a_{r}(t)$ and $%
b_{r}(t) $ are the new coefficients. \ The eigenvalue $P$ can be written as
: 
\begin{equation}
P=p+(n_{1}-n_{2})\hbar k,  \eqnum{10}  \label{10}
\end{equation}
where $p$ is chosen the same for both, ground and excited atomic levels. \
Integration over $q$ now yields to the following form: 
\begin{eqnarray}
\Psi &=&\left( 
\begin{array}{c}
0 \\ 
1
\end{array}
\right) \sum_{r=0}^{N}a_{r}(t)\left| r\right\rangle _{1}\left|
N-r\right\rangle _{2}\exp \left( \frac{i}{\hbar }(P-(N-2r)\hbar k)z\right) 
\eqnum{11}  \label{11} \\
&&+\left( 
\begin{array}{c}
1 \\ 
0
\end{array}
\right) \sum_{r=0}^{N-1}b_{r}(t)\left| r\right\rangle _{1}\left|
N-1-r\right\rangle _{2}\exp \left( \frac{i}{\hbar }(P-(N-1-2r)\hbar
k)z\right) .  \nonumber
\end{eqnarray}
This form is already eigen-one for the operator $\widehat{T}$ 
\begin{equation}
\widehat{T}\Psi =T\Psi ,  \eqnum{12}  \label{12}
\end{equation}
\begin{equation}
T=-\exp \left( \frac{i\pi }{\hbar k}\left( P-N\hbar k\right) \right) =-\exp
\left( \frac{i\pi }{\hbar k}p\right) .  \eqnum{13}  \label{13}
\end{equation}
\qquad

\bigskip Hence, the expression (\ref{11}) is the eigenfunction of operators $%
\widehat{N},\widehat{P}$ and $\widehat{T}$. \ Using(\ref{7}), (\ref{10}) and
the second equation in (\ref{13}), the latter three can be replaced by the
initial values $n_{1},n_{2}$ and $p.$

As it was previously stated, we are concentrating on the stationary states,
where 
\begin{equation}
a_{r}(t)=a_{r}\exp \left( -\frac{iEt}{\hbar }\right) ,\text{ }%
b_{r}(t)=b_{r}\exp \left( -\frac{iEt}{\hbar }\right) ,  \eqnum{14}
\label{14}
\end{equation}
$E$ being energy of the system.

After standard transformations the Schrodinger equation leads to following
set of tridiagonal recurrence algebraic equations: 
\begin{equation}
\left[ E-N\hbar \omega -\left( P-\left( N-2r\right) \hbar k\right) ^{2}/2M%
\right] a_{r}=\left( \hbar \Omega _{0}/2^{3/2}\right) \left[ \sqrt{r}b_{r-1}+%
\sqrt{N-r}b_{r}\right] ,  \eqnum{15a}  \label{15a}
\end{equation}
\begin{equation}
\left[ E+\hbar \varepsilon -N\hbar \omega -\left( P-\left( N-1-2r\right)
\hbar k\right) ^{2}/2M\right] b_{r}=\left( \hbar \Omega _{0}/2^{3/2}\right) %
\left[ \sqrt{r+1}a_{r+1}+\sqrt{N-r}a_{r}\right] ,  \eqnum{15b}  \label{15b}
\end{equation}
where $r=0,1,2,\cdot \cdot \cdot ,N$ and $\varepsilon =\omega -\omega _{0}$
is the atom-field frequency detuning.

Equations (\ref{15a}), (\ref{15b}), as well as equations (\ref{18a}) and (%
\ref{18b}) \ for the SW-case, are the basis of the remainder of this paper.
They determine the coefficients (probability amplitudes ) $a_{r}$, $b_{r}$
for ground and excited internal energy levels and the permitted values of
energy $E.$

The set of Eqs. (\ref{15a}), (\ref{15b}) has $2N+1$ solutions $E_{\nu
}(P,N), $ $\{a_{r}(\nu ,P,N),b_{r}(\nu ,P,N)\mid \nu =0,1,2,...,2N\}.$ \ $%
(N+1)$ of them may be interpreted as created adiabatically from initial
states with a ground-level atom, and the other $N$ ,as created respectively
with an excited-state atom. For simplicity we suppress $N,P$ and $\nu $ in
probability amplitude and energy notations.

Rather than pursuing the governing equations in this CPW-case we now turn to
another case, to SW-case. The respective Hamiltonian is \cite{4} 
\begin{equation}
H=\frac{1}{2M}\widehat{P}^{2}+\frac{1}{2}\left( 1+\sigma _{3}\right) \hbar
\omega _{0}+\hbar \omega a^{+}a+\frac{1}{4}\hbar \Omega _{0}(\sigma
_{+}a+\sigma _{-}a^{+})[\exp (ikz)+\exp (-ikz)],  \eqnum{16}  \label{16}
\end{equation}
where $a$ and $a^{+}$ are now SW-mode annihilation and creation operators,
obeying the same-type commutation relation $[a,a^{+}]=1$ as the running
modes. \ Note, that the coupling constant in this case is smaller by factor $%
2^{1/2}$ than in the former case.

The system ''atom + SW'' has three degrees of freedom, by one less than the
system with CPW. The reason is, that the momentum $\widehat{P}=\widehat{p},$%
and is not commutative with the Hamiltonian (\ref{16}). \ Proceeding the
same way as for (\ref{11}) and (\ref{15a}), ( \ref{15b}), we arrive to 
\begin{eqnarray}
\Psi  &=&\left( 
\begin{array}{c}
0 \\ 
1
\end{array}
\right) \left| N\right\rangle \sum_{l=-\infty }^{\infty }a_{l}(t)\exp \left( 
\frac{i}{\hbar }(p-2l\hbar k)z\right)   \eqnum{17}  \label{17} \\
&&+\left( 
\begin{array}{c}
1 \\ 
0
\end{array}
\right) \left| N-1\right\rangle \sum_{l=-\infty }^{\infty }b_{l}(t)\exp
\left( \frac{i}{\hbar }(p-(2l+1)\hbar k)z\right)   \nonumber
\end{eqnarray}
for wave function, and 
\begin{equation}
\left[ E-N\hbar \omega -\left( p-2l\hbar k\right) ^{2}/2M\right]
a_{l}=\left( \hbar \Omega _{0}/4\right) \sqrt{N}\left[ b_{l-1}+b_{l}\right] ,
\eqnum{18a}  \label{18a}
\end{equation}
\begin{equation}
\left[ E+\hbar \varepsilon -N\hbar \omega -\left( p-\left( 2l+1\right) \hbar
k\right) ^{2}/2M\right] b_{l}=\left( \hbar \Omega _{0}/4\right) \sqrt{N}%
\left[ a_{l+1}+a_{l}\right]   \eqnum{18b}  \label{18b}
\end{equation}
for probability amplitudes and energies. Here $l=0,\pm 1,\pm 2,...$ , and
the other notations are obvious. \ For adiabatically attained states 
\begin{equation}
N=\left\{ 
\begin{array}{c}
n\text{ for free ground level atom,} \\ 
n+1\text{ for free excited level atom. }
\end{array}
\right\} .  \eqnum{19}  \label{19}
\end{equation}

Besides the apparent similarities, the basic sets of equations, (\ref{15a}),
(\ref{15b}) and (\ref{18a}), (\ref{18b}), also have basic differences. \
While the first set is finite-size, the second one is infinite (even for $%
N=1,$ one-photon case). \ The next difference is in right-hand side
coefficients. \ They are varied ($r$-dependent) for the first, but are
constant for the second set. \ And the final difference, which already is
conceptual and deserves a peculiar attention, is that the total momentum $P$
in CPW-case, in contrast to its counterpart $p$ in SW case, cannot be
treated as a quasimomentum. \ This assertion immediately follows from the
fact, that the replacing $P\rightarrow P+s2\hbar k$, with the relabeling $%
r\rightarrow r-s$ in (\ref{15a}), (\ref{15b}), $s$ being an integer,
conserves the left-hand side coefficients invariant, while the right-hand
side coefficients get new values, leading thereby to other values for
eigenenergies and probability amplitudes. \ Reminding that the physical
reason of this difference is the participation of the momentum conservation
law, we arrive at a conclusion that $the$ $admission$ $of$ $the$ $%
quasimomentum$ $concept$ $in$ $space-periodic$ $\ systems$ $is$ $conditioned$
$by$ $the$ $failure$ $of$ $momentum$ $conservation$. \ $There$ $is$ $no$ $%
concept$ $of$ $quasimomentum$ $in$ $really$ $closed$ $systems$ $conserving$ $%
the$ $total$ $momentum.$

The main goal of the remainder of this section will be to illustrate the
behavior of both systems via the results of numerical calculations. While
the solutions of the CPW equations have been got exactly, to SW equations
has been applied suitably truncated matrix diagonalization method. \
Solutions depend, except the number $N$ of excitations (photons), upon the
relative size of four parameters: \ the resonance detuning $\varepsilon $,
vacuum Rabi-frequency $\Omega ,$ transverse kinetic energy detuning $Pk/2M$
and photon-induced kinetic detuning, which is the recoil energy $%
E_{r}=(2\hbar k)^{2}/2M$ in $\hbar $ units $.$ \ It is convenient to
introduce dimensionless parameters, scaling the energies by the recoil
energy $E_{r}$ and all momenta - by one-photon-reemission recoil momentum $%
2\hbar k.$ \ Then, as dimensionless parameter of interaction for both sets
of equations appears 
\begin{equation}
\xi =\hbar \Omega _{0}/2^{3/2}E_{r}.  \eqnum{20}  \label{20}
\end{equation}

The strength of coupling is essential already at $\xi \sim 1$. \ Let us for
illustration take the sodium's atom with $D_{2}$ $(3S_{1/2}-3P_{3/2})$ main
transition and estimate the volume $V$ of microcavity for which this is so,
more definitely: $\xi _{Na}=10.$ \ Using the explicit expressions for $%
\Omega _{0\text{ }}$and $E_{r}$ we arrive at 
\begin{equation}
\xi =\sqrt{\frac{3f}{2mV}}\frac{\left| e\right| M\lambda ^{2}}{2\pi
^{2}\hbar },  \eqnum{20a}  \label{20a}
\end{equation}
where $f$ and $\lambda $ are the oscillator strength and the wavelength of
optical transition, $\ m$ is the mass of electron. Inserting the values ($%
\lambda =589\ast 10^{-7}cm,$ $M=3.8\ast 10^{-23}g,$ $f=0.655$) we get $%
V=10^{-3}cm^{3}.$ Note, that for cesium atom, with the analogous transition $%
6S_{1/2}-6P_{3/2}$ and for the same volume of microcavity, the coupling
strength is more than one order stronger, $\xi _{Cs}\approx 135.$

Figures 1a and 1b show the energy spectrum as a function of momentum for CPW
and SW-cases respectively. \ In both cases $N=6,$ $V=10^{-3}cm^{3},$ $\hbar
\varepsilon /E_{r}=250,$ and the atom of $Na$ is chosen. The comparison of
figures clearly indicates the mentioned differences for energy spectrums:
first, the number of energy branches is finite ($2N+1=13$ $in$ $chosen$ $case
$) for CPW-case, whilst it is infinite for SW-case (are presented only some
lower branches); second, the CPW-spectrum is not repeated as $2\hbar k$
-wide zones, pointing out the failure of the quasimomentum concept in this
case. \ The Figure \ref{Fig.2} , plotted for the case $\ N=12$, shows, as it
was expected, approaching of the CPW-picture to the SW one. \ The flattening
of lower curves signifies the transition of these states into the
bounded-like states, for increased number of external photons. One can see,
that the gaps between the energy zones are essentially wider in CPW-case. \
Also is seen the well known regularity, that the widths of the gaps between
the $n^{th}$ and ($n+1)^{th}$ \ bands diminish with number $n$ \cite{11}.

Probability amplitude distributions are shown in Figures 3a, 3b and 4a, 4b,
respectively for $\ N=6$ and $\ N=12$ cases. As is seen from comparison of
the graphs, in the first pair ( $\ N=6$ full quantum range ) there is an
apparent discrepancy between CPW and SW-cases. \ In the second pair of
graphs ($N=12$) it is suppressed, but yet stays visible, even essential. \
The latter graphs also display the difference of momentum distributions for
bounded-like (lower-laying levels) and free-like (upper laying levels)
states.

The momentum distribution in bounded-like stationary states has a two-peak
form. \ The peaks retire from each other for increasing energies, gradually
being broader and losing in height. In free-like states , in contrast, the
distributions are centered at the nonperturbed values (i.e. at $l=0$ in
SW-case). To avoid the eventual misunderstanding of depicted graphs, it
needs to be noted that the momentum axes (abscissa) in these figures, as
well as in the following ones for probability amplitude distributions,
really contains only discrete values. Only for convenience of exhibition the
neighboring points in graphs have been merged via the straight lines

\section{Off-Resonance Approximation And Tavis-Cummings-Type Analytic
Solution}

In this section we examine the case, where the resonance detuning $%
\varepsilon $ is much larger than any other above mentioned frequency; $%
\Omega _{0,}P(p)k/2M$ and $\ E_{r}/\hbar .$ This case usually is regarded as
adiabatic elimination, or off-resonance approximation \cite{12}. \ In zero
order for $\varepsilon $ we obtain from (\ref{15b}) and (\ref{18b}) the
equations

$\bigskip $

\begin{equation}
b_{r}\simeq (\Omega _{0}/2^{3/2}\varepsilon )\left[ \sqrt{r+1}a_{r+1}+\sqrt{%
N-r}a_{r}\right] ,  \eqnum{21}  \label{21}
\end{equation}
$r=0,1,...,N$

\begin{equation}
b_{l}\simeq (\sqrt{N}\Omega _{0}/4\varepsilon )\left[ a_{l+1}+a_{l}\right] ,
\eqnum{22}  \label{22}
\end{equation}
$l=0,\pm 1,\pm 2,...$ respectively. \ This approximation, widely used in
problems without the kinetic energy operator in Hamiltonian, calls for some
comments here. \ The matter is, that for instance in SW-case, the term $%
(p-(2l+1)\hbar k)^{2}/2M$ already can exceed the any preliminarily given
value $\hbar \varepsilon $. But it has to be taken into account, that the
probability amplitudes, rather out of two-peaked range of distribution, are
extremely small and can be put out of consideration. \ Then we arrive at the
condition 
\begin{equation}
\left| \varepsilon \right| \gg \frac{\sqrt{N}\Omega _{0}}{4},  \eqnum{23}
\label{23}
\end{equation}
for implementation of the approximation, the same as the one without the
kinetic energy operator: detuning of the resonance must be much greater than
the optically induced inhomogeneous width of the transition.

By means of (\ref{21}) and (\ref{22}) \ the Eqs. (\ref{15a}) and (\ref{18a})
are being transformed into the recurrence equations only among the
ground-level atomic amplitudes: \ \ \ \ \ \ \ \ \ \ \ \ \ \ \ \ \ \ \ \ \ \
\ \ \ \ \ \ \ \ \ \ \ \ \ \ \ \ \ \ \ \ \ \ \ \ \ \ \ \ \ \ \ \ \ \ \ \ \ \
\ \ \ \ \ \ \ \ \ \ \ \ \ \ \ \ \ \ \ \ \ \ \ \ \ \ \ \ \ \ \ \ \ \ \ \ \ \
\ \ \ \ \ \ \ \ \ \ \ \ \ \ \ \ \ \ \ \ \ \ \ \ \ \ \ \ \ \ \ \ \ \ \ \ \ \
\ \ \ \ \ \ 
\begin{equation}
\lbrack E-N\hbar \omega -(P-(N-2r)\hbar k)^{2}/2M]a_{r}=\frac{\hbar \Omega
_{0}^{2}}{8\varepsilon }(\sqrt{r(N-r+1)}a_{r-1}+Na_{r}+\sqrt{(r+1)(N-r)}%
a_{r+1}),  \eqnum{24}  \label{24}
\end{equation}
\begin{equation}
\lbrack E-N\hbar \omega -(p-2l\hbar k)^{2}/2M]a_{l}=\frac{\hbar \Omega
_{0}^{2}N}{16\varepsilon }(a_{l-1}+2a_{l}+a_{l+1}).  \eqnum{25}  \label{25}
\end{equation}
This approximation preserves the nature of conjunctions between the
neighboring ground level amplitudes: $l\rightarrow l\pm 1$. \ Therefore, we
would expect that this approximation does not cause qualitative changes for
ground level stationary states. \ But it, as a matter of fact, strongly
depresses the excited-levels, created from the excited-level atomic states.
\ The probability amplitudes, calculated on the basis of Eqs. (\ref{24}) and
(\ref{25}), are plotted in Figures 5a and 5b. \ All the parameters are like
in Figures 4a and 4b and have to fulfill the condition (\ref{23}) ($\left|
\varepsilon \right| /(\sqrt{N}\Omega _{0}/4)\simeq 58$): Note that the
chosen value $\left| \varepsilon \right| $ is even larger than the natural
linewidth of envisioned optical transition.

Unfortunately, even approximated Eq. (\ref{24}) (or (\ref{25})) can not be
solved analytically due to $(2r\hbar k)^{2}/2M$ terms, quadric relative to
variable $r$ (or $l$). \ To proceed, we make a new supplementary
approximation, regarded to as a Tavis-Cummings-type approximation \cite{13}.
\ Presuming the $a_{r(l)}$-amplitude as a slowly varying function of $r$ $(l)
$ , a permissible step for $N\gg 1$ and if other parameters are out of Bragg
and Doppleron resonance conditions, this approximation envisages the
variable $r$ $(l)$ as a continuous one. \ It should be expected, of course,
to be a good approximation for $a_{r(l)}$ , when the momentum distribution
is rather dissipated.

Expanding the $a_{r\pm 1}$ in a Taylor series up to the second order, 
\begin{equation}
a_{r\pm 1}=a_{r}\pm \frac{da_{r}}{dr}+\frac{1}{2}\frac{d^{2}a_{r}}{dr^{2}}, 
\eqnum{26}  \label{26}
\end{equation}
and substituting this expression into the Eq. (\ref{24}) (CPW-case) \ we get
a second-order ordinary differential equation for the seeking probability
amplitudes and energy values 
\begin{equation}
\frac{d^{2}A(x)}{dx^{2}}-\frac{4}{N^{2}}\left( x-\frac{P}{2\hbar k}-r\right) 
\frac{dA_{x}}{dx}+\left\{ 4+\frac{4E_{r}}{U}\left[ \frac{E-N\hbar \omega }{%
E_{r}}-x^{2}\right] \right\} A(x)=0,  \eqnum{27}  \label{27}
\end{equation}
where $x=(P/2\hbar k)+r-(N/2),$ $U=-(\hbar \Omega _{0}^{2}/4\varepsilon )$
and the new amplitude $A(x)$ is introduced by means of a relation 
\begin{equation}
A(x)=\sqrt{r!(N-r)!}\text{ }a_{r}.  \eqnum{28}  \label{28}
\end{equation}

The respective equation in SW-case is 
\begin{equation}
\frac{d^{2}a(x)}{dx^{2}}+\left\{ 4+\frac{4E_{r}}{U}\left[ \frac{E-N\hbar
\omega }{E_{r}}-x^{2}\right] \right\} \text{ }a(x)=0,  \eqnum{29}  \label{29}
\end{equation}
where $x=(P/2\hbar k)+l,$ $U$ being the same as in the previous case. \ The
classical analog of $U$ is the average potential energy of atomic c.m. in
the light field.

Note, that the resulting Eqs. (\ref{27}) and (\ref{29}) are no longer
equivalent to the preliminary Eqs.(\ref{24}) and (\ref{25}) and some
specific quantum features of the atomic states may be lost (see in
particular the last paragraph of Section IV).

Let us first consider the more simple SW-case (Eq. (\ref{29})). \ We
restrict ourselves with the so called red-detuning case when $\varepsilon
\prec 0$ $(U\succ 0).$ \ Then the Eq.(\ref{29}) is the equation for
parabolic cylinder function $D_{\nu }(y)$ with a real argument. \ Its two
independent solutions are \cite{14} 
\begin{equation}
a_{1}=D_{\nu }(y),\text{ \ \ \ \ \ \ \ \ \ }a_{2}=D_{\nu }(-y),  \eqnum{30}
\label{30}
\end{equation}
where 
\begin{equation}
y=2\left( \frac{E_{r}}{U}\right) ^{1/4}x=2\left( \frac{E_{r}}{U}\right)
^{1/4}\left( \frac{p}{2\hbar k}+l\right) ,  \eqnum{31}  \label{31}
\end{equation}
\begin{equation}
\nu =-\frac{1}{2}+\frac{E-N\hbar \omega +U}{\sqrt{E_{r}U}},  \eqnum{32}
\label{32}
\end{equation}

The requirement for $D_{\nu }(\pm y)$ to be limited for any values of $y$
(variable $l$) leads to a condition $\nu =n=0,1,2,...$. Then in accordance
with (\ref{32}), the energy gets only discrete values and is determined by
the formula 
\begin{equation}
E_{n}=N\hbar \omega -U+\sqrt{E_{r}U}\left( n+\frac{1}{2}\right) .  \eqnum{33}
\label{33}
\end{equation}
Simultaneously both relevant parabolic cylinder functions turn to a Hermite
polynomial $H_{n}(y/\sqrt{2})$ and as a sequence 
\begin{equation}
a_{n,l}=\frac{C_{n}}{\sqrt{2^{n}n!}}H_{n}\left( \sqrt{2}\left( \frac{E_{r}}{U%
}\right) ^{1/4}\left( \frac{p}{2\hbar k}+l\right) \right) \exp \left[ -\sqrt{%
\frac{E_{r}}{U}}\left( \frac{p}{2\hbar k}+l\right) ^{2}\right] ,  \eqnum{34}
\label{34}
\end{equation}
where $C_{n}$ is a normalizing constant. \ It is worth to note that the size
of quantization in (\ref{33}) exactly copies the known frequency of time
oscillations in the problem of atomic diffraction in the classical
standing-wave field, taking into account the c.m. motion along the field
direction.

In CPW-case, in distinction with SW-one, we arrive at 
\begin{equation}
E_{n}=N\hbar \omega +\frac{P^{2}}{2M}\frac{1/N^{2}}{1/N^{2}+E_{r}/U}-U+\frac{%
\hbar \Omega _{0}^{2}}{8\varepsilon }+\sqrt{\frac{\hbar ^{2}\Omega _{0}^{4}}{%
16\varepsilon ^{2}}+UE_{r}}\left( n+\frac{1}{2}\right) ,  \eqnum{35}
\label{35}
\end{equation}
and 
\begin{equation}
a_{n,r}=C_{n}\frac{\exp \left[ \frac{1}{N}\left( \frac{N}{2}-r\right) ^{2}%
\right] }{\sqrt{r!(N-r)!}}H_{n}\left( \frac{y(r)}{\sqrt{2}}\right) \exp %
\left[ -\frac{y^{2}(r)}{4}\right] ,  \eqnum{36}  \label{36}
\end{equation}
where 
\begin{equation}
y\left( r\right) =2\left( \frac{1}{N^{2}}+\frac{E_{r}}{U}\right)
^{1/4}\left( \frac{P}{2\hbar k}+r-\frac{N}{2}-\frac{1}{1/N^{2}+E_{r}/U}\frac{%
1}{N^{2}}\frac{P}{2\hbar k}\right) .  \eqnum{37}  \label{37}
\end{equation}

CPW energy spectrum (\ref{35}) has two distinguishing features in respect to
SW spectrum (\ref{33}). \ The first is, that the total momentum $P$ enters
into the expression of energy. \ This means that the atomic c.m. motion is
not bounded in stationary states in general ($P\neq 0$), but keeps some
one-sided mean velocity. \ States become bounded only in the limit $%
N^{2}\rightarrow \infty ,$ that is, in classical field limit. \ As a scale
of this limit is claimed the ratio $U/E_{r}.$ Taking into account the
definition of $U$ (just after Eq.(\ref{27})), the mentioned condition may be
written in the form 
\begin{equation}
N\gg \frac{\hbar \Omega _{0}^{2}}{4\left| \varepsilon \right| E_{r}}, 
\eqnum{38}  \label{38}
\end{equation}
provided, of course, the condition (\ref{23}). \ For $V=10^{-3}cm^{3}$
cavity the condition (\ref{38}) is satisfied at $N\succcurlyeq 10$ for
sodium atoms and at $N\succcurlyeq 50$ for cesium atoms.

The second difference of (\ref{35}) relative to (\ref{33}) is the size of
quantization, emerging by the additional term $\hbar ^{2}\Omega
_{0}^{4}/16\varepsilon ^{2}$ in square root expression. \ It has negligible
influence in the viewed off-resonance approximation. \ 

Evident differences occur in probability amplitudes. \ It is convenient to
image the expression in (\ref{36}) as a product of two parts; first,
including the Hermite polynomial and second, the ratio of exponential to the
square root of factorials. \ First part coincides with the (\ref{34}), but
is shifted relative it by 
\begin{equation}
\frac{1}{1/N^{2}+E_{r}/U}\frac{1}{N^{2}}\frac{P}{2\hbar k}.  \eqnum{39}
\label{39}
\end{equation}
This shift has the same coefficient at $P/2\hbar k$ as the kinetic energy
term in (\ref{35}), and is not small in quantum optics range ($N\sim 10$ for
off-resonance approximation). \ The second part, i.e. the additional ratio,
is a symmetric function relative to point $r=N/2$, the form of which is
presented in Fig.\ref{Fig.6} (for some values of $N$). \ As is seen, it can
be viewed as unity in central region and decreasing near the edges of the
definition range. \ Hence, the amplitude distribution in CPW-case, being
product of two symmetric by itself, but shifted with respect to each other
functions, is not, already, a symmetric one, as it was in CPW-case. \ It is
symmetric only if $P=0$, but even in this case, the CPW distribution is
somewhat narrower than the SW distribution. \ In analogy with energy
spectrum, all differences disappear in the limit $N^{2}\rightarrow \infty ,$
where the counterpropagating light waves appear to a far detuned two-level
atom as a sinusoidal potential, just similar to the standing light wave.

Figures 7a and 7b show the probability distributions calculated by means of
formulas (\ref{36}) and (\ref{34}) respectively. \ All the parameters are
like in Figures 5a and 5b. Comparison of \ Fig.7a with Fig.7b shows a small,
almost imperceptible narrowing in each CPW-''mountain range''. The mutual
comparison of Figures 7a and 7b with Figures 5a and 5b corroborates the
Tavis-Cummings-type approximation to be an acceptable tool in the range of
bounded states. But it totally puts out of consideration all free or
quasi-free states. In Fig.8a we present a similar to Fig.7a distribution,
calculated for conditions more precisely satisfying Tavis-Cummings-type
approximation ($N=24$). \ Its coincidence with the original, see Fig.8b, is
of course better. \ However, it is to be memorized that the rising of $N$
drifts the theory to the semiclassical domain of interactions.

\section{\protect\bigskip Formation Of Stationary States}

Let us return to the theme of stationary states in the cases under
consideration. Our objective is to clarify the ideas about the quantum
contents of interaction mechanisms in SW and CPW-cases, to indicate the
differences between them and already from this point, to understand the
formation of stationary states from initially free states. \ To present
them, we will first consider the SW-case. \ The variation of atomic c.m.
motion in this case can be imagined as a diffraction of atomic matter waves
on ''frozen'' immovable object: the standing wave grating. \ This is a fully
classical-like phenomenon, provided by the matter-wave representation for
atomic c.m. motion. \ The picture, that the scattering is a result of
absorption and reemission of SW-photons should be regarded as a defective,
since it does not answer to the question how the atomic momentum is changing
due to these zero-momentum photons. \ This obstacle is overcame in the
theory by the fact that the probability amplitude distribution $a_{l}$ with $%
l=0,\pm 1,\pm 2,...$ , as well as the amplitude distribution $b_{l},$
represent only the atomic states and do not concern to photons. \ 

To realize the adiabatic evolution of the system from initially free state
to the stationary coupling one, we assume the parameter $\Omega _{0}$ of the
interaction to be a slow function of time, replace $E$ by $i\hbar (d/dt)+E$
in (\ref{25}) \ (in off-resonance approximation) and view the $a_{l}$%
-amplitudes and the parameter $E$ as a slowly varying in time quantities. \
After a phase transformation 
\begin{equation}
\alpha _{l}=a_{l}\exp \left[ -\frac{i}{\hbar }\left( E-N\hbar \omega
-(p+2l\hbar k)^{2}/2M-\frac{\hbar \Omega _{0}^{2}N}{8\varepsilon }\right) t%
\right] ,  \eqnum{40}  \label{40}
\end{equation}
we arrive at the equation 
\begin{eqnarray}
\frac{d}{dt}\left| \alpha _{l}\right| ^{2} &=&-\frac{N\Omega _{0}^{2}}{%
8\varepsilon }(\cos (\nu _{l-1/2}t)%
%TCIMACRO{\func{Im}}%
%BeginExpansion
\mathop{\rm Im}%
%EndExpansion
(\alpha _{l}\alpha _{l-1}^{\ast })+\cos (\nu _{l+1/2}t)%
%TCIMACRO{\func{Im}}%
%BeginExpansion
\mathop{\rm Im}%
%EndExpansion
(\alpha _{l}\alpha _{l+1}^{\ast })+  \eqnum{41}  \label{41} \\
&&\sin (\nu _{l-1/2}t)%
%TCIMACRO{\func{Re}}%
%BeginExpansion
\mathop{\rm Re}%
%EndExpansion
(\alpha _{l}\alpha _{l-1}^{\ast })+\sin (\nu _{l+1/2}t)%
%TCIMACRO{\func{Re}}%
%BeginExpansion
\mathop{\rm Re}%
%EndExpansion
(\alpha _{l}\alpha _{l+1}^{\ast }))  \nonumber
\end{eqnarray}
for density matrix elements, which are more preferable for the putted
objective. \ Here $\nu _{l}\equiv 2\hbar k(p+2l\hbar k)/M$ is the analog of
Doppler frequency shift.

The equation (\ref{41}) is, of course, too complicated to be inspected in
general form, but we may consider the initial stage of interaction, taking
in addition $\alpha _{l}(t=0)=\delta _{l,l_{0}}.$ \ Then $\alpha _{l\pm 1}$
are fully imaginary with $\ 
%TCIMACRO{\func{Im}}%
%BeginExpansion
\mathop{\rm Im}%
%EndExpansion
\alpha _{l\pm 1}$ $\succ 0$ (for $\varepsilon \prec 0$) and the relative
rate of depletion of the state with $l=l_{0}-1$ and $l=l_{0}+1$ is given by
relative values of $\cos (\nu _{l_{0}-1/2}t)$ and $\cos (\nu _{l_{0}+1/2}t)$
respectively. \ If $\ p+2l_{0}\hbar k\gtrless 0,$ then $\cos (\nu
_{l_{0}-1/2}t)\gtrless \cos (\nu _{l_{0}+1/2}t)$ and less modulo momentum
states are being filled more rapidly than more modulo momentum states. \ As
a result, the mean atomic momentum decreases gradually, gathering for
asymptotically long times about the zero value. \ In attained stationary
state, the mean atomic momentum approaches zero. \ Note, that this is a
stimulated analog of well known Doppler cooling (slowing) of red-detuned
atoms in the standing wave. \ The case of CPW we will start (and mainly
restrict ourselves) by the immovable atom model (about commutative set of
operators see in Appendix), since it gives an opportunity to present the
additional mechanism \ for stationary state formation. \ Putting $M=\infty $
in Eqs. (\ref{15a}) and (\ref{15b}) and substituting the second into the
first we readily arrive at equation 
\begin{equation}
\frac{d}{dt}\left| \alpha _{r}\right| ^{2}=-\frac{\hbar \Omega _{0}^{2}}{%
4[E+\hbar \varepsilon -N\hbar \omega ]}\left[ \sqrt{r(N-r+1)}%
%TCIMACRO{\func{Im}}%
%BeginExpansion
\mathop{\rm Im}%
%EndExpansion
(\alpha _{r}\alpha _{r-1}^{\ast })+\sqrt{(r+1)(N-r)}%
%TCIMACRO{\func{Im}}%
%BeginExpansion
\mathop{\rm Im}%
%EndExpansion
(\alpha _{r}\alpha _{r+1}^{\ast })\right]   \eqnum{42}  \label{42}
\end{equation}
for density matrix elements, where 
\begin{equation}
\alpha _{r}=a_{r}\exp \left[ -\frac{i}{\hbar }\left( E-N\hbar \omega -\frac{%
\hbar ^{2}\Omega _{0}^{2}N}{8[E+\hbar \varepsilon -N\hbar \omega ]}\right) t%
\right] .  \eqnum{43}  \label{43}
\end{equation}
Here, in contrast to SW-case, each atomic c.m. state is single-bonded with
the pair of photon numbers in CPW. \ The mathematical response of this
connection is that the $a_{r}($and $b_{r})$ amplitudes are the
representation of both atomic c.m. and photonic states simultaneously. \ The
difference between the rates of $r\rightarrow r-1$ and $r\rightarrow r+1$
transitions now is determined by square roots $\sqrt{r(N-r+1)}$ and $\sqrt{%
(r+1)(N-r)}.$ \ The former term represents the process of reemission of a
photon from the first travelling wave, containing preliminary $r$ photons,
into the second one, with respectively $N-r$ photons, concomitantly altering
the atomic momentum in $2\hbar k.$ \ The second term represents the inverse
process, i.e. the reemission of a photon from the second wave into the first
wave, accompanying it with a $2\hbar k$ atomic momentum change. \ The
physical reason of differences between the mentioned rates is solely the
bosonic nature of photons. \ Direct comparison shows the dominance of
reemission of photons from ''more photon'' wave into the ''less photon''
wave, which finally results in equalization of mean photon numbers in both
counterpropagating travelling waves \cite{15}. \ Hence, in attained
stationary states the mean field momentum approaches zero.

Thus, both analyzed cases possess diverse mechanisms of approaching to
stationary state, but the final results are similar in the sense, that since
one of the coupling subsystems, atom or field, is ''immovable'' or
''frozen'', the mean momentum of the other subsystem unavoidably approaches
zero. \ Speaking in fancy form, the ''immovable'' part of the system takes
on itself the ''surplus'' of the momentum from the other subsystem. \ A
clear and expected result. \ In CPW-case with a moving atom, however, this
result is not the general case. \ The reason is the conservation of system's
momentum. \ In this case the system evolves under action of both mechanisms.
\ First of them, the diffraction, tends to suppress the mean atomic
momentum, whilst second one, the bosonic nature, in contrary, tends to
suppress the difference between the mean numbers of counterpropagating
travelling waves. \ Since they are not possible simultaneously (except the
case $P=0$), in the course of time they mature the stationary states with
definite, asymmetric in general, statistical distributions, where these
tendencies are mutually compatible. The role of bosonic mechanism however is
essential in the quantum-optic domain ($N\curlyeqprec 10$) and gradually
diminishes out of it.

To conclude this section, we would like to make a notice about one
unavoidable peculiarity of Tavis-Cummings-type approximation, with respect
to adiabaticity of interaction. \ Taking the photon momentum as a continuous
variable, we have suppressed the size of its quantum up to zero (as small as
possible). \ Thereby we make feasible the arbitrary small variations in the
energy of atomic c.m. motion. \ As a consequence, any as small as possible
variation of the field intensity (more concrete of parameter $\Omega _{0}$)
should cause changes in atomic c.m. motion and breaks down the adiabatic
approximation. \ Because of this, the approximate solutions (\ref{34}) and (%
\ref{36}), in contrast to exact Eqs. (\ref{15a}), (\ref{15b}) or (\ref{18a}%
), (\ref{18b}), cannot be adiabatically related to the free system's state.
\ Continuous quantum transition in atomic c.m. motion completely ''washes
out'' the information about the initial state and its energy. \ Really, the
switching off of interaction ($\Omega _{0}\rightarrow 0$) in Tavis
Cummings-type solutions leads to only one $E(\Omega _{0})=N\hbar \omega $
value of energy and to one corresponding probability amplitude with $%
r=\left( N/2)+(P/2\hbar k\right) $\ photons in the first travelling wave and
the rest $N-r=\left( N/2)-(P/2\hbar k\right) $ photons in the second
travelling wave. \ It is necessary, however, to clearly realize that the
parameters $n_{1},n_{2}$ and $p$ preserve their physical essence as
parameters of initial state; they are determined by the relations (\ref{7}),
(\ref{10}) and (\ref{13}) (in CPW-case) and are $\Omega _{0}$-independent. \
Simply, the Tavis-Cummings-type approximation interrupts any, even
adiabatic, connection of the stationary state with the initial one. \ The
parameters $n_{1},n_{2}$ and $p$ can be used in this approximation as
externally (additionally) given ones, and in case of necessity can be
corroborated by means of exact equations.

\section{Bragg And Doppleron Resonances}

These resonances, Bragg and Doppleron, have been identified for an atom
moving in a standing wave \cite{16} and have got implementations in atom
optics and laser cooling \cite{17}. \ Here we would consider only the limit
of small intensities, focusing on the role of multiphoton recoil energy.

Some preliminary words, for reminding, how these resonances are being
extracted from equations. \ Thinking of the limit $\Omega _{0}\rightarrow 0$
as a leading one to the initial state, one should put a corresponding
condition on the initial state amplitudes. \ For example, if $n_{1\text{ }}$%
and $n_{2}$ photons in CPW and the atom is on the ground level, then $%
a_{r}=\delta _{r,n_{1}},b_{r}=0.$ \ As a consequence, we get from (\ref{15a}%
) the value of the free system energy, 
\begin{equation}
E(\Omega _{0}=0)=p^{2}/2M+(n_{1}+n_{2})\hbar \omega .  \eqnum{44}  \label{44}
\end{equation}
Here the atomic momentum $p$ is arbitrary. When in contrary, the limit $%
\Omega _{0}\rightarrow 0$ is thought of as a transition to the final free
state, we may only confirm the equality of energy to its (\ref{44}) free
value. \ It follows from the conservation of energy. \ Also taking into
account the relations (\ref{7}) and (\ref{10}), from Eqs. (\ref{15a}), (\ref
{15b}) we directly arrive at 
\begin{equation}
\left( n_{1}-r\right) (p+(-n_{1}+r)\hbar k)a_{r}=0,  \eqnum{45}  \label{45}
\end{equation}
or 
\begin{equation}
(\omega -\omega _{0}-\left( 2n_{1}-2r-1\right) kp/M-\left(
2n_{1}-2r-1\right) ^{2}\hbar k^{2}/2M)b_{r}=0.  \eqnum{46}  \label{46}
\end{equation}

The solution with $r=n_{1}$ and respectively $a_{n_{1}}=1,a_{n\neq
n_{1}}=0,b_{r}=0,$ returns the system into the initial state. \ The next
solution with 
\begin{equation}
p=\left( -n_{1}+r\right) \hbar k  \eqnum{47}  \label{47}
\end{equation}
gives a sole nonzero value for $a_{r=n_{1}+p/\hbar k}$ ground-level
amplitude. \ Corresponding to this case final atomic momentum (see the wave
function (\ref{11})) $p_{f}=P+\left( N-2r\right) \hbar k=p+2\left(
n_{1}-r\right) \hbar k=-p,$ which is the condition for Bragg scattering.

The third possible solution of (\ref{45}) and (\ref{46}) occurs at 
\begin{equation}
\omega -\omega _{0}=\left( 2n_{1}-2r-1\right) kp/M+\left( 2n_{1}-2r-1\right)
^{2}\hbar k^{2}/2M  \eqnum{48}  \label{48}
\end{equation}
and has a sole nonzero solution for the corresponding excited-level
amplitude $b_{r}$. \ This is the case of Doppleron-resonance scattering,
when the atom leaves the zone of interaction on excited energy level \cite
{18}.

Let us stop our attention on Doppleron-resonance condition (\ref{48}) in
order to examine the behavior of final momentum $p_{f}$ versus the initial
momentum $p$. \ To this end the condition (\ref{48}) may be viewed as a
dependance of $r$ on $p$ and then put into the excited-atom final momentum
expression $p_{f}=P+\left( N-1-2r\right) \hbar k=p+\left( 2n_{1}-1-2r\right)
\hbar k.$ \ Computation gives 
\begin{equation}
p_{f}=-p\sqrt{1+2M\hbar \left( \omega -\omega _{0}\right) /p^{2}}, 
\eqnum{49}  \label{49}
\end{equation}
where the atomic initial momentum satisfies the Doppleron resonance
condition (\ref{48}). For a fixed frequency value $\omega $ the allowed
values of initial momentum $p$ constitute a discrete set, corresponding to
the integer values of $\left( 2n_{1}-2r-1\right) $ in (\ref{48}).

Graphs of (\ref{49}) are plotted in Fig.\ref{Fig.9}. \ For comparison we
also plot the graphs without the recoil energy term (the last one in (\ref
{48})). \ As we can see, the presence of recoil energy dramatically changes
the character of $p_{f}(p)$ dependance. \ The first distinctive feature is
in splitting of possible values of $p_{f}$ for every $p$ into two branches,
herewith converging the high-order resonances ($p^{2}/2M\gg \hbar \left|
\omega -\omega _{0}\right| $) $\ $to the Bragg resonances ($p_{f}=-p)$ in
the additional branch. \ The second difference is the lack of resonances in
small atomic momentum range ($p^{2}/2M\prec \hbar \left| \omega -\omega
_{0}\right| $) and connected with it, the convergence of $p_{f}$ \ to zero
at the limiting point with $p^{2}/2M=\hbar \left| \omega -\omega _{0}\right|
.$ \ It means that the inherent possibilities of the Doppleron-cooling
really are more than it follows from the theory, not including the recoil
energy term.

\section{Conclusions}

We analyze the stationary states of JCM with SW and CPW fields, treating the
external, as well as the internal, atomic coordinates quantum-mechanically.
\ The inspection shows essential deviations between SW- and CPW-cases in
quantum optics range ($N\prec 20$), which first of all have qualitative
character. \ As a such matter can be noted the deterioration of the concept
of quasimomentum, the asymmetry and narrowing of momentum distribution, the
coincidence of photonic and atomic momentum distributions in CPW-case.

Since the physical reason of the mentioned differences is the existence of
the momentum conservation, it can be concluded that the modern cavity QED an
the theory of measurement, being developed on the base of quantized
SW-cavities, in fact have only dealt with the conservation of energy. \ The
cavity QED and related problems for ring cavities, where the momentum
conservation exists too and has to alternate the state-pattern in
''atom+field'' system, need a performance, yet. \ Moreover, the experimental
confirmation just of CPW-case regularities, stationary or not, may be
adopted as the direct nonmediated evidence for the photon momentum to be
quantized.

Reexamination of kinematic resonances shows that due to recoil energy term
the Doppleron-resonance splits into two branches and , which is more
important, can not take place in the range of small incident kinetic
energies (where $\left| p\right| \prec \sqrt{2M\hbar \left| \omega -\omega
_{0}\right| }$).

This work was supported by ISTC Grant A - 215 - 99.

\section{Appendix}

The case of immovable atom ($\widehat{p}^{2}/2M=0$) is a separate one for
stationary state problem and its analysis is not logical on the basis of
Eqs. (\ref{15a}), (\ref{15b})$.$ The\ reason\ for\ such\ a\ separation is
that the $\widehat{P}$ and $\widehat{T}$ operators loose the inherent
physical contents and must be omitted. Instead of them enters a new,
commutative with $\widehat{H}$ and\ $\widehat{N}$ photonic operator \ $\ $%
\begin{equation}
\widehat{\theta }%
=(a_{1}^{+}a_{1}+a_{2}^{+}a_{2}-a_{1}^{+}a_{2}-a_{2}^{+}a_{1})/2N, 
\eqnum{A1}  \label{A1}
\end{equation}
which has the eigenstates $\theta _{m}=1-m/N,$ $m=0,1,2,\cdot \cdot \cdot
,N. $

It is referred to as an operator of interference, in the sense that its
eigenvalues play the same role in the quantized-field picture as the term $%
cos^{2}kz$ in the classical field picture of interaction \cite{16}.

\begin{center}
\bigskip

\begin{figure}[tbp]
\caption{{}Dispersion curves of the atomic c.m. quantum motion in
counterpropagating waves (a) and standing wave (b) quantized cavity fields
with $N=6$ photons. \ The chosen parameters are $\hbar \protect\varepsilon
/E_{r}=-250,$ $\protect\xi =\hbar \Omega _{0}/2^{3/2}E_{r}=1.75$. \ The
latter corresponds to $3S_{1/2}-3P_{3/2}$ optical transition of sodium atom
located in a $V=10^{-3}cm^{3}$ quantum cavity. \ $n=1,$ $2,$ $...$ on curves
labels the number of energy zone. The lowest states, up to $n=4,$ are shown.
\ CPW-curves (a), in distinction to SW-ones, aren't periodic and, as a
result, the concept of quasimomentum can not be introduced for them. \ The
energy distance between the $n=1$ and $n=2$ curves, that is the width of
first energy band, is much wider in CPW-case than in SW-case. }
\label{Fig.1}
\end{figure}
\begin{figure}[tbp]
\caption{{}Dispersion curves of the atomic c.m. quantum motion in
counterpropagating waves (a) and standing wave (b) quantized cavity fields
with $N=12$ photons. The other parameters are like in Fig.1. \ The wider
widths of energy bands in CPW-case still is aparent. \ The second band
between $n=2$ and $n=3$ zones is clearly seen in CPW-case, whilst in SW-case
it is not.}
\label{Fig.2}
\end{figure}
\begin{figure}[tbp]
\caption{{}Distribution of the atomic c.m. momentum states over the energy
spectrum. \ The probabilities are presented on vertical axes. \ The momenta
are on the frontal axes and are presented in $p_{r}=2\hbar k$ units. \ The
energies are presented without photonic field energy in $E_{r}=(2\hbar
k)^{2}/2M$ recoil energy units. \ The total momentum $P$ (quasimomentum $p$)
of ''atom +cavity field'' system is chosen zero. The other parameters are
like in Fig.1. \ The values of energy are quantized and depicted by dark
lines on both graphs. \ The spectrums of momenta are quantized too, but for
sake of simplicity of exhibition we connect the neighboring points in both
momentum and energy directions. \ The lower-laying energy levels are created
from ground-level free atomic states, and the upper-laying ones are from
excited-level free atomic states. \ The energy spectrum of CPW-case (a) is
not truncated, whilst the spectrum of SW-case (b) in reality is infinite and
we exhibit only some truncated parts for both, lower-laying and upper-laying
families.\ Finally the distributions in a form of expanding hills pertain to
quasi-bounded states; the others are localized near the zero-momentum values
and represent the quasi-free states. }
\label{Fig.3}
\end{figure}
\begin{figure}[tbp]
\caption{{}Distribution of the atomic c.m. momentum states over the energy
spectrum for $N=12$ photons in cavities. \ The parameters are like in the
previous figure. \ The momentum distributions in bounded states are somewhat
wider than in previous $N=6$ case. \ The distinction of CPW-case (a) from
SW-case (b) is essential yet. }
\label{Fig.4}
\end{figure}
\begin{figure}[tbp]
\caption{{}The same momentum distribution as in Fig.4, but calculated in
off-resonance approximation. Excited-state energy levels are escaped, of
course, while the ground-state energy levels and c.m. momentum distributions
practically coincide with the corresponding results in Fig.4. }
\label{Fig.5}
\end{figure}
\begin{figure}[tbp]
\caption{{}The function-factor, which determines the form distinction of
c.m. momentum distribution in CPW-case with respect to SW-case (in
Tavis-Cummings approximation), as function of number $r$ of photons in one
of the counterpropagating waves.}
\label{Fig.6}
\end{figure}

\begin{figure}[tbp]
\caption{{}Atomic c.m. momentum distribution in Tavis-Cummings approximation
for CPW (a) - and SW (b) -cases. \ The parameters are taken from Fig.5. \
The continuous spectrum absent in this approximation and the difference
between CPW- and SW-cases is notable only for upper-levels.}
\label{Fig.7}
\end{figure}

\begin{figure}[tbp]
\caption{{}Momentum distribution for atomic c.m. quantum motion in
off-resonance and Tavis-Cummings approximations with $\ N=24$ photons in
cavities. As was expected the coincidence is better (with the lower bounded
states) than in the previous $\ N=12$ case, but is not appropriate for
quantitative calculations.}
\label{Fig.8}
\end{figure}
\begin{figure}[tbp]
\caption{{}Final atomic momentum as a function of initial (incident)
momentum for Doppleron-resonance conditions with (solid line) and without
(dashed line) recoil energy term. \ The graphs are depicted as continuous
ones (instead of discrete point sequences) only for the sake of simplicity.}
\label{Fig.9}
\end{figure}
\end{center}

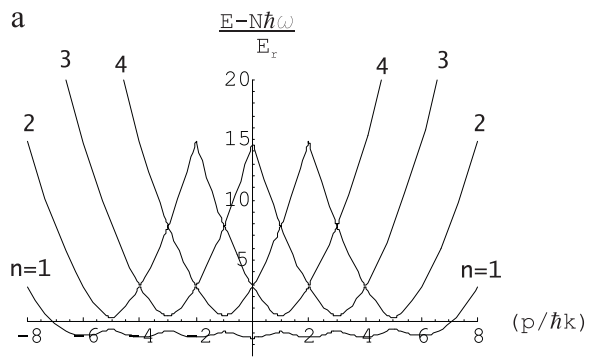
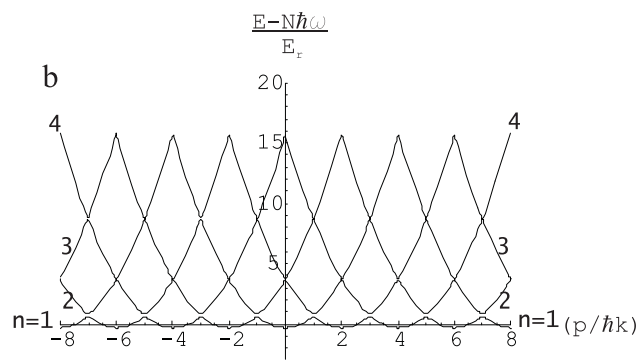
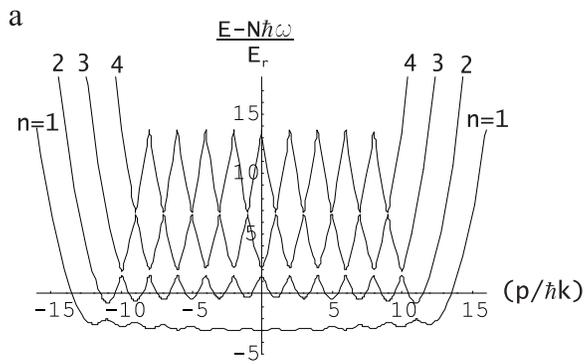
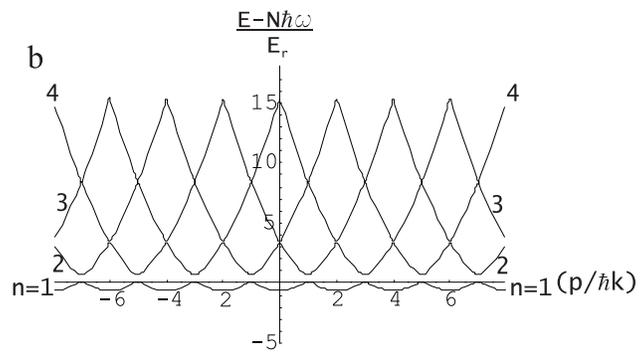
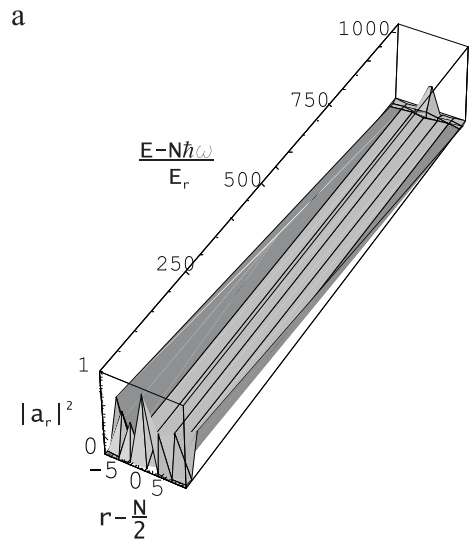
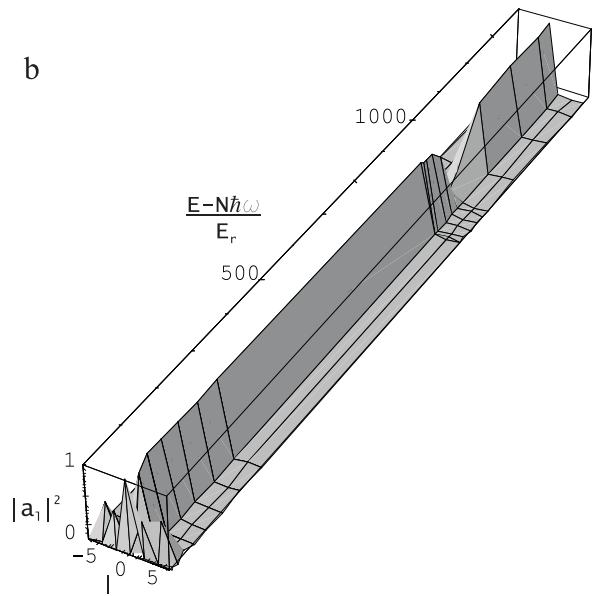
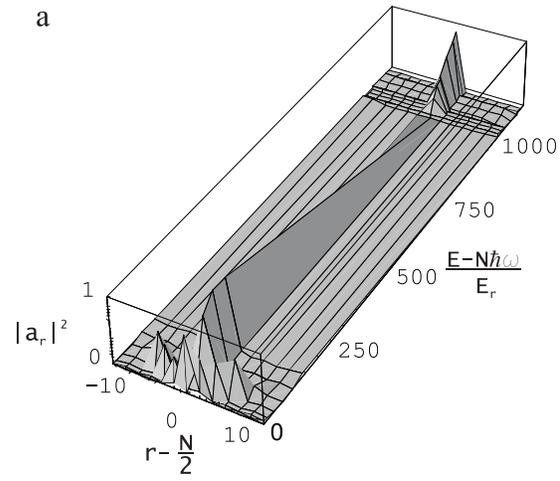
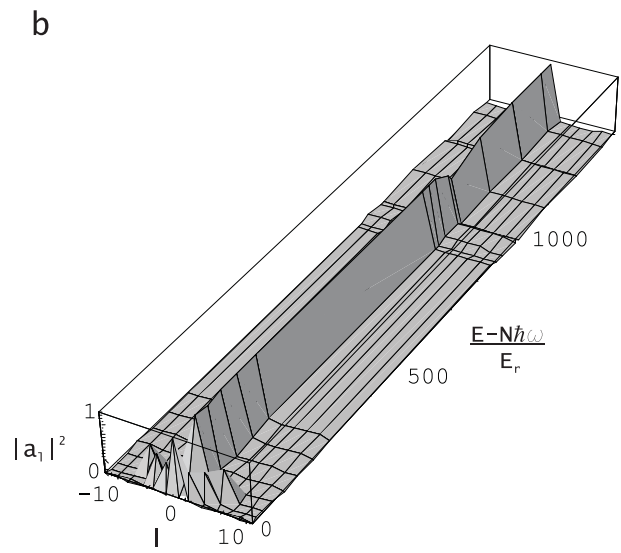
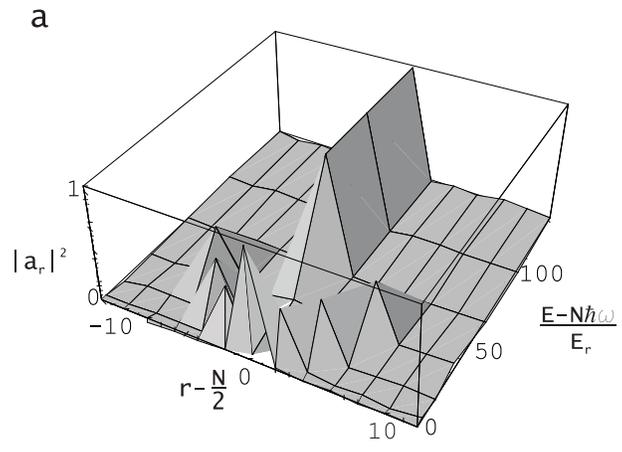
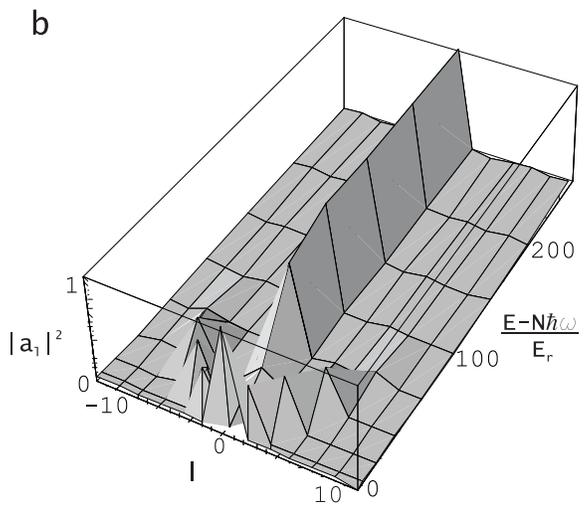
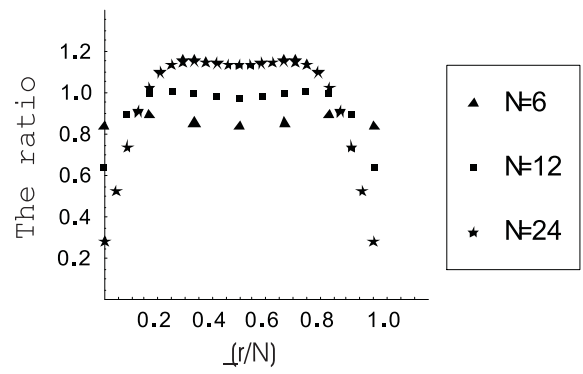
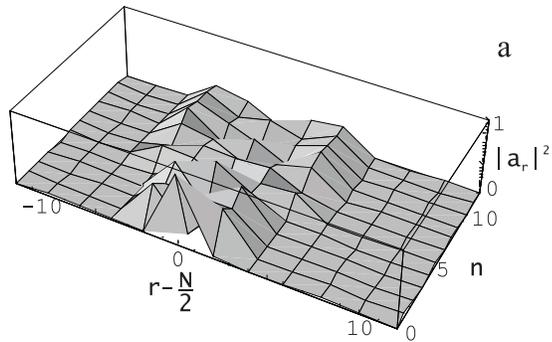
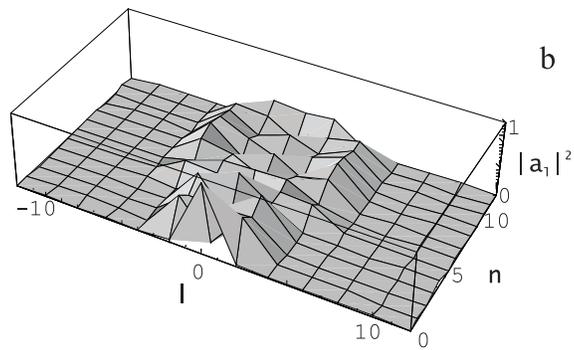
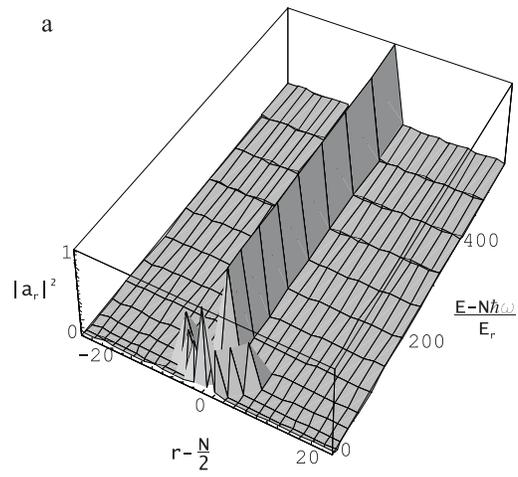
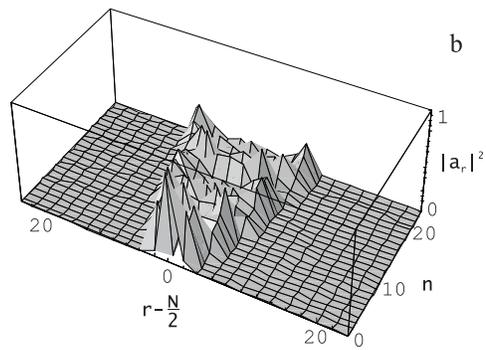
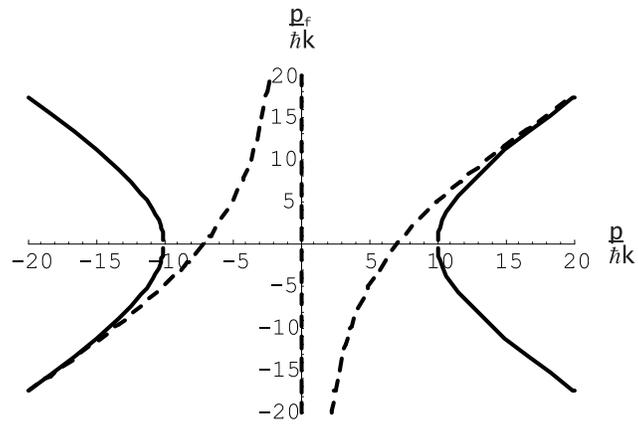
\end{document}